# Subjective Information Measure and Rate Fidelity Theory

Chenguang Lu[①]
Independent Researcher,
Survival99@hotmail.com

*Abstract*—Using fish-covering model, this paper intuitively explains how to extend Hartley's information formula to the generalized information formula step by step for measuring subjective information: metrical information (such as conveyed by thermometers), sensory information (such as conveyed by color vision), and semantic information (such as conveyed by weather forecasts). The pivotal step is to differentiate condition probability and logical condition probability of a message. The paper illustrates the rationality of the formula, discusses the coherence of the generalized information formula and Popper's knowledge evolution theory. For optimizing data compression, the paper discusses rate-of-limiting-errors and its similarity to complexity-distortion based on Kolmogorov's complexity theory, and improves the rate-distortion theory into the rate-fidelity theory by replacing Shannon's distortion with subjective mutual information. It is proved that both the rate-distortion function and the rate-fidelity function are equivalent to a rate-of-limiting-errors function with a group of fuzzy sets as limiting condition, and can be expressed by a formula of generalized mutual information for lossy coding, or by a formula of generalized entropy for lossless coding. By analyzing the rate-fidelity function related to visual discrimination and digitized bits of pixels of images, the paper concludes that subjective information is less than or equal to objective (Shannon's) information; there is an optimal matching point at which two kinds of information are equal; the matching information increases with visual discrimination (defined by confusing probability) rising; for given visual discrimination, too high resolution of images or too much objective information is wasteful.

## I. INTRODUCTION

To measure sensory information and semantic information, I set up a generalized information theory thirteen years ago [4-8] and published a monograph focusing on this theory in 1993 [5]. But, my researches are still rarely known by English researchers of information theory. Recently, I read some papers about complexity distortion theory [2], [9] based on Kolmogorov's complexity theory. I found that, actually, I had discussed complexity-distortion function and proved that the generalized entropy in my theory was just such a function, and had concluded that the complexity-distortion function with size-unequal fuzzy error-limiting balls could be expressed by a formula of generalized mutual information. I also found that some researchers did some efforts [9] similar to mine for improving Shannon's rate-distortion theory.

This paper first explains how to extend Hartley's information formula to the generalized information formula, and then discusses the generalized mutual information and some questions related to Popper's theory, complexity distortion theory, and rate-distortion theory.

## II. HARTLEY'S INFORMATION FORMULA AND A STORY OF COVERING FISH

Hartley's information formula is [3]

$$I=\log N, \qquad (1)$$

where $I$ denotes the information conveyed by the occurrence of one of $N$ events with equal probability. If a message $y$ tells that uncertain extension changes from $N_1$ to $N_2$, then information conveyed by $y$ is

$$I_r=I_1-I_2=\log N_1-\log N_2=\log(N_1/N_2). \qquad (2)$$

We call (2) relative information formula. Before discussing its properties, Let's hear a story about covering fish with fish covers.

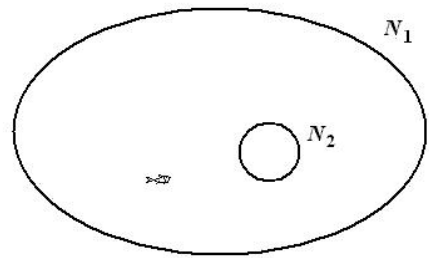

Figure 1: Fish-covering model for relative information $I_r$.

Fish covers are made of bamboo. A fish cover looks like a hemisphere with a round hole at top for human hand to catch fish. Fish covers are suitable for catching fish in adlittoral pond. When I was a teenage, after watching peasants catch fish with fish covers, I decided to do the same thing. I found a basket with a hole at bottom and followed those peasants to catch fish. Fortunately, I successfully caught some fish, but no so much as the peasants did. Then I compared my smaller basket with much bigger fish cover to get the following conclusions.

The fish cover is bigger so that covering fish is easier; yet, catching fish with hands is more difficult. If a fish cover is big enough to cover the pond, it must be able to cover fish. However, this huge fish cover is useless because catching fish with hands is as difficult as without fish cover. When one uses the basket or smaller fish cover to cover fish, though covering fish is more difficult, catching fish with hands is much easier.

An uncertain event is alike a fish with random position in a pond. Let a sentence $y$="Fish is covered"; $y$ will convey information about the position of fish. Let $N_1$ be the area of the pond, $N_2$ be the area covered by the fish cover, then information conveyed by $y$ is $I_r=\log(N_1/N_2)$. The smaller $N_2$ is than $N_1$, the bigger the information amount is. This just reflects the advantage of the basket. If $N_2 = N_1$, then $I=0$. This just tells us that covering fish is meaningless if the cover is as big as the pond. The above formula cannot tell the advantage of fish covers (covering fish with less failure) in comparison with the basket because in the classical information theory, there seems a hypothesis that the failure of covering fish never happens. The generalized information formula introduced bellow will contain "possible failure of covering fish".

### III. IMPROVING THE FISH-COVERING FORMULA WITH PROBABILITY

Hartley's information formula requires $N$ events with equal probability $P=1/N$. Yet, the probabilities of events are unequal in general. For example, the fish stays in deep water in bigger probability and in shallow water in smaller probability. In these cases, we need to replace $1/N$ with probability $P$ so that
$$I = \log(1/P) \quad (3)$$
and
$$I_r = \log(P_2/P_1). \quad (4)$$

### IV. RELATIVE INFORMATION FORMULA WITH A SET AS CONDITION

Let $X$ denote the random variable taking values from set $A=\{x_1, x_2, \ldots\}$ of events, $Y$ denote the random variable taking values from set $B=\{y_1, y_2, \ldots\}$ of sentences or messages. For each $y_j$, there is a subset $A_j$ of $A$ and $y_j = "x_i \in A_j"$, which can be cursorily understood as "Fish $x_i$ is in cover $A_j$". Then $P_1$ above becomes $P(x_i)$, $P_2$ becomes $P(x_i | x_i \in A_j)$. We simply denote $P(x_i | x_i \in A_j)$ by $P(x_i | A_j)$, which is conditional probability with a set as condition. Hence, the above relative information formula becomes
$$I(x_i; y_j) = \log[P(x_i | A_j)/P(x_i)]. \quad (5)$$

For convenience, we call this formula as the fish-covering information formula.

Note that the most important thing is generally $P(x_i|A_j) \neq P(x_i|y_j)$, because
$$P(x_i|y_j) = P(x_i|"x_i \in A_j") = P(x_i|"x_i \in A_j" \text{ is reported});$$
yet,
$$P(x_i|A_j) = P(x_i|x_i \in A_j) = P(x_i|"x_i \in A_j" \text{ is true}),$$
where $y_j$ may be an incorrect reading datum, a wrong message, or a lie, yet, $x_i \in A_j$ means that $y_j$ must be correct. If they are always equal, then formula (5) will become classical information formula
$$I(x_i; y_j) = \log[P(x_i | y_j)/P(x_i)], \quad (6)$$
whose average is just Shannon mutual information [11].

### V. BAYESIAN FORMULA FOR THE FISH-COVERING INFORMATION FORMULA

Let the feature function of set $A_j$ be $Q(A_j|x_i) \in \{0,1\}$. According to Bayesian formula, there is
$$P(x_i|A_j) = Q(A_j|x_i)P(x_i)/Q(A_j), \quad (7)$$

where $Q(A_j) = \sum_i P(x_i)Q(A_j|x_i)$. From (5) and (7), we have
$$I(x_i; y_j) = \log\frac{P(x_i|A_j)}{P(x_i)} = \log\frac{Q(A_j|x_i)}{Q(A_j)}, \quad (8)$$

which (illustrated by Figure 2) is the transition from classical information formula to generalized information formula.

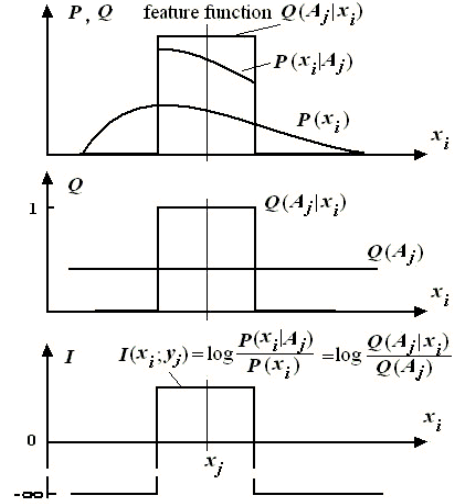

Figure 2: Illustration of fish-covering information formula related to Bayesian formula.

Let us use a thermometer to explain how to use the fish-covering information formula to measure metrical information.

The reading datum of a thermometer may be considered to be reporting sentence $y_j \in B=\{y_1, y_2, \ldots\}$, and real temperature as the fish position $x_i \in A=\{x_1, x_2, \ldots\}$. Let $y_j = "x_i \in A_j"$, and $A_j = [x_j - \triangle x, x_j + \triangle x]$ according to the resolution of the thermometer and eyes' visual discrimination. Hence, we can use the fish-covering information formula to measure thermometric information.

### VI. GENERALIZED INFORMATION FORMULA WITH A FUZZY SET AS CONDITION

Information conveyed by a reading datum of thermometer and information conveyed by a forecast "The rainfall will be about 10 mm" are the same in essence. Using a clear set as condition as above is not good enough because the information amount should change with $x_i$ continuously. We wish that the bigger the error (i.e. $x_i - x_j$), the less the information. Now, using a fuzzy set to replace the clear set as condition, we can realize this purpose (see Figure 4).

Now, we consider $y_j$ to be sentence "$X$ is $x_j$" (or say $y_j = \hat{x}_j$). For a fuzzy set $A_j$ whose feature function $Q(A_j|x_i)$ takes value from [0, 1] and $Q(A_j|x_i)$ can be considered to be confusing probability of $x_i$ with $x_j$. If $i=j$, then the confusing probability reaches its maximum 1.

Actually, the confusing probability $Q(A_j|x_i)$ is only different parlance of the membership grade of $x_i$ in fuzzy set $A_j$ or the logical probability of proposition $y_j(x_i)$. There is
$Q(A_j|x_i)$ = feature function of $A_j$
   = confusing probability or similarity of $x_i$ with $x_j$
   = membership grade of $x_i$ in $A_j$
   = logical probability or creditability of proposition $y_j(x_i)$

The discrimination of human sense organs, such as visual discrimination for gray levels of pixels of images, can also be described by confusing probability functions. In these cases, a sensation can be considered to be a reading datum $y_j = \hat{x}_j$ of the thermometer. The visual discrimination function of $x_j$ is $Q(A_j|x_i)$, $i=1, 2,...$ where $A_j$ is a fuzzy set containing all $x_i$ that are confused with $x_j$. We may use the statistic of random clear sets to obtain this function [13].

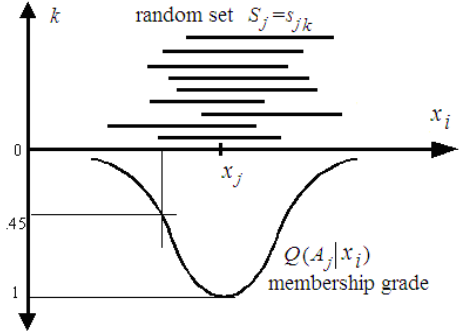

Figure 3: Confusing probability function from clear sets.

First we do many times experiments to get the clear confusing sets $s_{jk}$, $k=1, 2…n$, by putting $x_j$ on one side of a screen and changing $x_i$ on another side of the screen for eyes to discern. And then we calculate

$$Q(A_j | x_i) = \frac{1}{n}\sum_k Q(s_{jk} | x_i), \quad (9)$$

Now, by replacing a clear set with a fuzzy set as condition, we get the generalized information formula:

$$I(x_i; y_j) = \log\frac{P(x_i | A_j)}{P(x_i)} = \log\frac{Q(A_j | x_i)}{Q(A_j)}. \quad (10)$$

It looks the same as the fish-covering information formula (8), but $Q(A_j|x_i) \in [0,1]$ instead of $Q(A_j|x_i) \in \{0,1\}$. And also, this formula allows wrong reading data or messages, bad forecasts, or lies which convey negative information. The generalized information formula can be understood as fish-covering information formula with fuzzy cover. Because of fuzziness, generally, the amount of negative information is finite. The property of the formula can be illustrated by Figure 4.

Figure 4 tells us that when a reading datum or a sensation $y_j = \hat{x}_j$ is provided, the bigger the difference of $x_i$ from $x_j$, the less the information; and the less the $Q(A_j)$, the bigger the absolute value of information. From this formula, we can conclude that information amount not only depends on the correctness of reflection, but also depends on the precision of reflection.

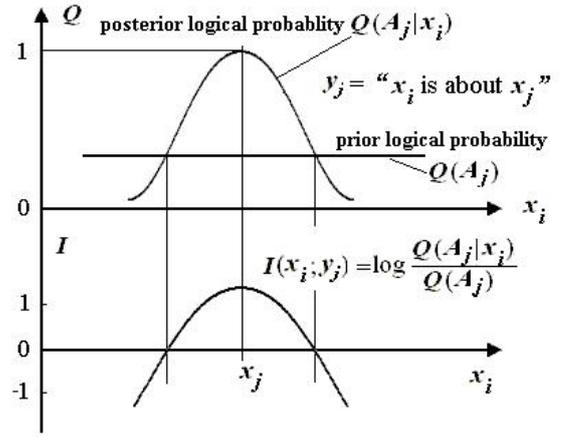

Figure 4: Generalized information formula for measuring metrical information, sensory information, and number-forecasting information.

VII. COHERENCE OF THE SEMANTIC INFORMATION MEASURE AND POPPER'S CRITERION OF ADVANCE OF KNOWLEDGE

The generalized information formula can also be used to measure semantic information in general, such as information from weather forecast "Tomorrow will be rainy or heavy rainy". We may assume that for any proposition $y_j$, there is a Plato's idea $x_j$ that makes $Q(A_j|x_j)=1$. The idea $x_j$ is probably not in $A_j$. Hence, any logical condition probability $Q(A_j|x_i)$ can be considered to be the confusing probability of $x_i$ with the idea $x_j$.

From my view-point, forecasting information is more general information in comparison with descriptive information. If a forecast is always correct, then the forecasting information will become descriptive information.

About the criterion of advance of scientific theory, philosopher Karl Popper wrote:

*"The criterion of relative potential satisfactoriness... characterizes as preferable the theory which tell us more; that is to say, the theory which contains the greater amount of empirical information or content; which is logically strong; which has the greater explanatory and predictive power; and which can therefore be more severely tested by comparing predicted facts with observations. In short, we prefer an interesting, daring, and highly informative theory to a trivial one."* ( in [10], pp. 250)

Clearly, Popper used information as the criterion to value the advance of scientific theories. According to Popper's theory, the more easily a proposition is falsified logically and the more it can go through facts (in my words, the less the prior logical probability $Q(A_j)$ is, and the bigger the posterior logical probability $Q(A_j|x_i)$ is ), the more information it conveys and the more meaningful it is. Contrarily, a proposition that can not be falsified logically (in my words, $Q(A_j|x_i) = Q(A_j) = 1$) conveys no information and is insignificant in science. Obviously, the generalized information measure is very coherent with Popper's information criterion; the generalized information formula functions as a bride between Shannon's information theory and Popper's knowledge evolution theory.

## VIII. GENERALIZED KULLBACK'S INFORMATION AND GENERALIZED MUTUAL INFORMATION

Calculating the average of $I(x_i; y_j)$ in (10), we have generalized Fullback's information formula for given $y_j$:

$$I(X; y_j) = \sum_i P(x_i | y_j) \log \frac{P(x_i | A_j)}{P(x_i)}. \quad (11)$$

Actually, the probabilities on the right of the log should be prior probabilities or logical probabilities, the probability on the left of the log should be posterior probability. Since now we differentiate two kinds of probabilities and use $Q(.)$ for those probabilities after log. Hence the above formula becomes

$$I(X; y_j) = \sum_i P(x_i | y_j) \log \frac{Q(x_i | A_j)}{Q(x_i)}. \quad (12)$$

We can prove that as $Q(X|A_j) = P(X|A_j)$, which means subjective probability forecasts conforms to objective statistic, $I(X; y_j)$ reaches its maximum. The more different the $Q(X)$ is from $P(X|A_j)$, which means that the facts are more unexpected, the bigger the $I(X; y_j)$ is. This formula also conforms to Popper's theory.

Further, we have generalized mutual information formula

$$\begin{aligned} I(X;X) &= \sum_j P(y_j) I(X; y_j) \\ &= \sum_i P(x_i, y_j) \log[Q(x_i | A_j)/Q(x_i)] \\ &= H(X) - H(X | Y) = H(Y) - H(Y | Y) \end{aligned} \quad (13)$$

where

$$H(X) = -\sum_i P(x_i) \log Q(x_j) \quad (14)$$

$$H(X | Y) = -\sum_j \sum_i P(x_i, y_i) \log Q(x_i | A_j) \quad (15)$$

$$H(Y) = -\sum_j P(y_j) \log Q(A_j) \quad (16)$$

$$H(Y | X) = -\sum_j \sum_i P(x_i, y_i) \log Q(A_j | x_i) \quad (17)$$

I call $H(X)$ forecasting entropy, which reflects the average coding length when we economically encode $X$ according to $Q(X)$ while real source is $P(X)$, and reaches its minimum as $Q(X) = P(X)$. I call $H(X|Y)$ posterior forecasting entropy, call $H(Y)$ generalized entropy, and call $H(Y|X)$ generalized condition entropy or fuzzy entropy [6].

I think that the generalized information is subjective information and Shannon information is objective information. If two weather forecasters always provide opposite forecasts and one is always correct and another is always incorrect. They convey the same objective information, but the different subjective information. If $Q(X) = P(X)$ and $Q(X|A_j) = P(X|y_j)$ for each $j$, which means subjective forecasts conform to objective facts, then the subjective mutual information equals objective mutual information.

## IX. RATE-OF-LIMITING-ERRORS AND ITS RELATION TO COMPLEXITY-DISTORTION

In [5], I defined rate-of-limiting-errors, which is similar to complexity distortion [2]. The difference is that the error-limiting condition for rate-of-limiting-errors is a group of sets or fuzzy sets $A_J = \{A_1, A_2...\}$ instead of a group of balls with the same size and clear boundaries for complexity distortion.

We know that the color space of digital images is visually ununiform and human eyes' discrimination is fuzzy. So, in some cases, such as coding for digital images, using size-unequal balls or fuzzy balls as limiting condition will be more reasonable.

Assume $P(Y)$ is a source; encode $Y$ into $X$; allow $y_j$ is encoded into any $x_j$ in clear set $A_j$, $j=1, 2...$; then the minimum of Shannon mutual information for different $P(X|Y)$ is defined as rate-of-limiting-errors $R(A_J)$.

Interestingly, it can be proved that $R(A_J)$ is just equal to the generalized entropy $H(Y)$ [5]. To realize this rate, there must be $P(X/y_j) = Q(X/A_j)$ for each $j$. Furthermore, when the limiting sets are fuzzy, i.e. $P(X/y_j) \leq Q(X/A_j)$ for each $j$ as $Q(A_j|\text{xi}) < 1$, there is

$$R(A_J) = \sum_j \sum_i P(y_j) Q(x_i | A_j) \log \frac{Q(A_j | x_i)}{Q(A_i)} \quad (18)$$

To realize this rate, there must be $P(X)=Q(X)$ and $P(X/y_j) = Q(X/A_j)$ for each $j$ so that Shannon's mutual information equals the generalized mutual information.

Now, from the view-point of the complexity distortion theory, the generalized entropy $H(Y)$ is just prior complexity, the fuzzy entropy $H(Y|X)$ is just the posterior complexity, and $I(X;Y)$ is the reduced complexity.

## X. RATE FIDELITY THEORY: REFORMED RATE DISTORTION THEORY

Actually, Shannon mentioned fidelity criterion for lossy coding before. He used the distortion as the criterion for optimizing lossy coding because the fidelity criterion is hard to be formulated. However, distortion is not a good criterion in most cases.

How do we value a person? We value him according to not only his errors but also his contributions. For this reason, I replace the error function $d_{ij}=d(x_i, y_j)$ with generalized information $I_{ij}= I(x_i; y_j)$ and distortion $d(X, Y)$ with generalized mutual information $I(X; Y)$ as criterion to search the minimum of Shannon mutual information $I_s(X; Y)$ for given $P(X)=Q(X)$ and the lower limit $G$ of $I(X; Y)$. I call this criterion the fidelity criterion, call the minimum the rate-fidelity function $R(G)$, and call the reformed theory the rate fidelity theory.

In a way similar to that in the classical information theory [1], we can obtain the expression of function $R(G)$ with parameter $s$:

$$\begin{aligned} G(s) &= \sum_j \sum_i P(x_i) P(y_j) \exp(sI_{ij}) \lambda_i I_{ij} \\ R(s) &= sG(s) + \sum_j \sum_i P(x_i) \log \lambda_i \end{aligned} \quad (19)$$

where $s=dR/dG$ indicates the slope of function $R(G)$ ( see Figure 5) and $\lambda_i = 1/\sum_j P(y_j) \exp(sI_{ij})$.

We define a group of sets $B_I = \{B_1, B_2...\}$, where $B_1, B_2...$ are subset of $B=\{y_1, y_2,...\}$, by fuzzy feature function

$$Q(B_i | y_j) = \exp(sI_{ij})/m = [Q(A_j | x_i)/Q(A_i)]^s / m \quad (20)$$

where $m$ is the maximum of $\exp(sI_{ij})$; then from (19) and (20) we have

$$R(G) = \sum_j \sum_i P(x_i) P(y_j | B_i) \log \frac{Q(B_i | y_j)}{Q(B_i)} \quad (21)$$
$$= R(B_I).$$

This function is just the rate-of-limiting-errors with a group of fuzzy sets $B_I=\{B_1, B_2…\}$ as limiting condition while coding $X$ in $A$ into $Y$ in $B$. From this formula, we can find there is profound relationship between rate-of-limiting-errors and rate-fidelity (or rate-distortion). In the above formulas, if we replace $I_{ij}$ with $d_{ij}=d(x_i, y_j)$, (21) is also tenable. So, actually rate-distortion function can be expressed by a formula of generalized mutual information.

In [7], I defined information value $V$ by the increment of growing speed of fund because of information, and suggested to use the information value as criterion to optimize communication in some cases to get function rate-value $R(V)$, which is also meaningful.

## XI. Rate-fidelity Function for Optimizing Image Communication

Now let's examine the relationships among subjectively visual information, visual discrimination, and objective information. For simplicity, we consider the information provided by different gray levels of pixels of images (see [4] for details).

Let the gray level of digitized pixel be a source and the gray level is $x_i=i$, $i=0, 1... b =2^k -1$ with normal probability distribution whose expectation$=b/2$ and standard deviation$= b/8$. Assume that after decoding, the pixel also has gray level $y_j=j=0, 1... b$; the perception caused by $y_j$ is also denoted by $y_j$; and discrimination function or confusing probability function of $x_j$ is

$$Q(A_j | X) = \exp[-(X - j)^2 /(2d^2)] \quad (22)$$

where $d$ is discrimination parameter. The smaller the $d$, the higher the discrimination.

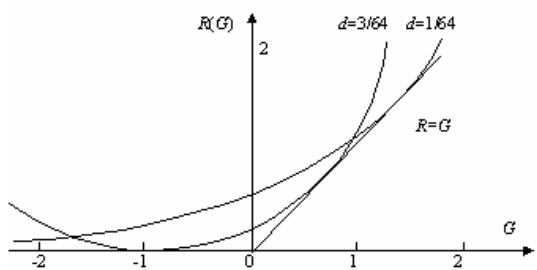

Figure 5. Relationship between $d$ and $R(G)$ for $b$=63.

Figure 5 indicates that when $R$=0, $G$<0, which means that if a coded image has nothing to do with the original image, we still believe it reflects the original image, then the information will be negative. When $G$=-2, $R$>0, which means that certain objective information is necessary when one uses lies to deceive his or her enemy to some extent; or say, lies against facts are more terrible than lies according to nothing. The each line of function $R(G)$ is tangent with the line $R=G$, which means there is a matching point at which objective information is equal to subjective information, and the higher the discrimination (the less the $d$), the bigger the matching information amount. The slope of $R(G)$ becomes bigger and bigger with $G$ increasing. This tells us for given discrimination, it is limited to increase subjective information.

Figure 6 tells us that for given discrimination, there exists the optimal digitized-bit $k'$ so that the matching value of $G$ and $R$ reaches the maximum. If $k<k'$, the matching information increases with $k$; if $k>k'$, the matching information no longer increases with $k$. This means that too high resolution of images is unnecessary or uneconomical for given visual discrimination.

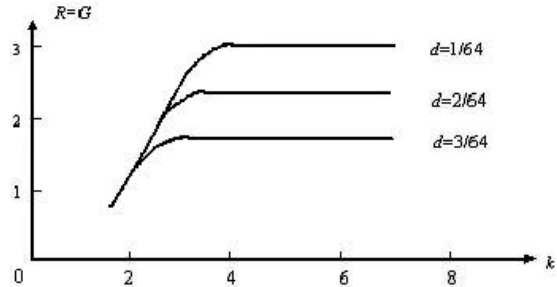

Figure 6: Relationship between matching value of $R$ with $G$, discrimination parameter $d$, and digitized bit $k$.

## XII. Conclusions

This paper has deduced generalized information formula for measuring subjective information by replacing condition probability with logical condition probability, and improved the rate-distortion theory into the rate fidelity theory by replacing Shannon distortion with subjective mutual information. It has also discussed the rate-fidelity function related to visual discrimination and digitized grades of images, and gotten some meaningful results.

References

[1] T. Berger, *Rate Distortion Theory*, Englewood Cliffs, N.J.: Prentice-Hall, 1971.
[2] M. S. Daby and E. Alexandros, "Complexity Distortion Theory", *IEEE Tran. On Information Theory,* Vol. 49, No. 3, 604-609, 2003.
[3] R. V. L. Hartley, "Transmission of information", *Bell System Technical Journal,* 7 , 535, 1928.
[4] C. Lu, "Coherence between the generalized mutual information formula and Popper's theory of scientific evolution"(in Chinese), *J. of Changsha University*, No.2, 41-46, 1991.
[5] C. Lu, *A Generalized Information Theory* (in Chinese), China Science and Technology University Press, 1993. see http://survivor99.com/lcg/books/GIT
[6] C. Lu, "Coding meaning of generalized entropy and generalized mutual information" (in Chinese), *J. of China Institute of Communications*, Vol.15, No.6, 38-44, 1995.
[7] C. Lu, *Portfolio's Entropy Theory and Information Value*, (in Chinese), China Science and Technology University Press, 1997
[8] C. Lu, "A generalization of Shannon's information theory", *Int. J. of General Systems*, Vol. 28, No.6, 453-490, 1999. see http://survivor99.com/lcg/english/information/GIT/index.htm
[9] G. Peter and P. Vitanyi, Shannon information and Kolmogorov complexity, *IEEE Tran. On Information Theory,* submitted, http://homepages.cwi.nl/~paulv/papers/info.pdf
[10] K. Popper, *Conjectures and Refutations—the Growth of Scientific Knowledge*, Routledge, London and New York, 2002.
[11] C. E. Shannon, "A mathematical theory of communication", *Bell System Technical Journal,* Vol. *27*, pt. I, pp. 379-429; pt. II, pp. 623-656, 1948.
[12] P. Z. Wang, *Fuzzy Sets and Random Sets Shadow* (in Chinese), Beijing Normal University Press, 1985.